\documentclass[aps,prl,twocolumn,tightenlines,amssymb,aps,nobibnotes,superscriptaddress,showpacs,balancelastpage,floatfix]{revtex4-1}
\usepackage[utf8]{inputenc}
\usepackage[utf8]{inputenc}
\usepackage{amsmath,revsymb,amssymb}
\usepackage{graphicx}
\usepackage[colorinlistoftodos]{todonotes}
\usepackage{xcolor}
\usepackage{tabularx}
\usepackage{multirow}
\usepackage{booktabs}
\usepackage{framed}
\usepackage{hyperref}
\usepackage[normalem]{ulem}
\hypersetup{
 colorlinks,
 linkcolor={red},
 citecolor={blue},
 urlcolor={blue}
}
\renewcommand{\emph}{\textit}

\begin{document}
    
    \title{High-precision spectroscopy of $^{20}$O benchmarking \textit{ab-initio} calculations in light nuclei}

\author{I.~Zanon}
	\affiliation{INFN Laboratori Nazionali di Legnaro, Legnaro, Italy.}
    \affiliation{Dipartimento di Fisica e Scienze della Terra, Universit\`a di Ferrara, Ferrara, Italy.}
\author{E.~Cl\'ement}
	\affiliation{Grand Acc\'el\'erateur National d'Ions Lourds (GANIL), CEA/DRF-CNRS/IN2P3, Caen, France}
\author{A.~Goasduff}
	\affiliation{INFN Laboratori Nazionali di Legnaro, Legnaro, Italy.}
\author{J. Men\'endez}
    \affiliation{Department of Quantum Physics and Astrophysics and Institute of Cosmos Sciences, University of Barcelona, Spain}
\author{T.~Miyagi}
	\affiliation{Technische Universit\"at Darmstadt, Department of Physics, Darmstadt, Germany}
\affiliation{ExtreMe Matter Institute, GSI Helmholtzzentrum f\"ur Schwerionenforschung GmbH, Darmstadt, Germany}
\affiliation{Max-Planck-Institut f\"ur Kernphysik, Heidelberg, Germany}
\author{M.~Assi\'e}
  \affiliation{Universit\'e Paris-Saclay, CNRS/IN2P3, IJCLab, 91405 Orsay, France}
\author{M.~Ciema\l{}a}
	 \affiliation{IFJ PAN, Krakow, Poland.}
\author{F.~Flavigny}
     \affiliation{Universit\'e de Caen Normandie, ENSICAEN, CNRS/IN2P3, LPC Caen UMR6534, F-14000 Caen, France.}
\author{A.~Lemasson}
	\affiliation{Grand Acc\'el\'erateur National d'Ions Lourds (GANIL), CEA/DRF-CNRS/IN2P3, Caen, France}
\author{A.~Matta}
     \affiliation{Universit\'e de Caen Normandie, ENSICAEN, CNRS/IN2P3, LPC Caen UMR6534, F-14000 Caen, France.}
\author{D.~Ramos}
	\affiliation{Grand Acc\'el\'erateur National d'Ions Lourds (GANIL), CEA/DRF-CNRS/IN2P3, Caen, France}
 \author{M.~Rejmund}
    \affiliation{Grand Acc\'el\'erateur National d'Ions Lourds (GANIL), CEA/DRF-CNRS/IN2P3, Caen, France}
\author{L.~Achouri}
     \affiliation{Universit\'e de Caen Normandie, ENSICAEN, CNRS/IN2P3, LPC Caen UMR6534, F-14000 Caen, France.}
\author{D.~Ackermann}
	\affiliation{Grand Acc\'el\'erateur National d'Ions Lourds (GANIL), CEA/DRF-CNRS/IN2P3, Caen, France}
\author{D.~Barrientos}
  \affiliation{CERN, CH-1211 Geneva 23, Switzerland}
\author{D.~Beaumel}
  \affiliation{Universit\'e Paris-Saclay, CNRS/IN2P3, IJCLab, 91405 Orsay, France}
\author{G.~Benzoni}
  \affiliation{INFN Sezione di Milano, I-20133 Milano, Italy}
\author{A.J.~Boston}
  \affiliation{Oliver Lodge Laboratory, The University of Liverpool, Liverpool, UK.}
\author{H.C.~Boston}
  \affiliation{Oliver Lodge Laboratory, The University of Liverpool, Liverpool, UK.}
\author{S.~Bottoni}
	 \affiliation{Dipartimento di Fisica, Universit\`a di Milano, Milano, Italy}
    \affiliation{INFN Sezione di Milano, I-20133 Milano, Italy}
\author{A.~Bracco}
  \affiliation{INFN Sezione di Milano, I-20133 Milano, Italy}
  \affiliation{Dipartimento di Fisica, Universit\`a di Milano, Milano, Italy}
\author{D.~Brugnara}
	\affiliation{INFN Laboratori Nazionali di Legnaro, Legnaro, Italy.}
	\affiliation{Dipartimento di Fisica, Universit\`a di Padova, Padova, Italy.}
\author{G.~de~France}
	\affiliation{Grand Acc\'el\'erateur National d'Ions Lourds (GANIL), CEA/DRF-CNRS/IN2P3, Caen, France}
\author{N.~de~Sereville}
  \affiliation{Universit\'e Paris-Saclay, CNRS/IN2P3, IJCLab, 91405 Orsay, France}
\author{F.~Delaunay}    
     \affiliation{Universit\'e de Caen Normandie, ENSICAEN, CNRS/IN2P3, LPC Caen UMR6534, F-14000 Caen, France.}
\author{P.~Desesquelles}
  \affiliation{Universit\'e Paris-Saclay, CNRS/IN2P3, IJCLab, 91405 Orsay, France}
\author{F.~Didierjean}
	 \affiliation{Universit\'e de Strasbourg, IPHC, Strasbourg, France.}
\author{C.~Domingo-Prato}
  \affiliation{Instituto de Fisica Corpuscolar, CSIC-Universidad de Valencia, E-46071 Valencia, Spain.}
\author{J.~Dudouet}
	 \affiliation{Université de Lyon, Université Lyon-1, CNRS/IN2P3, UMR5822, IP2I, F-69622 Villeurbanne Cedex, France}
\author{J.~Eberth}
  \affiliation{Institut für Kernphysik, Universität zu Köln, Zülpicher Str. 77, D-50937 Köln, Germany}
\author{D.~Fern\'andez}
	 \affiliation{IGFAE and Dpt. de Física de Partículas, Univ. of Santiago de Compostela, Santiago de Compostela, Spain}
\author{C.~Foug\`eres}
	\affiliation{Grand Acc\'el\'erateur National d'Ions Lourds (GANIL), CEA/DRF-CNRS/IN2P3, Caen, France}
\author{A.~Gadea}
  \affiliation{Instituto de Fisica Corpuscolar, CSIC-Universidad de Valencia, E-46071 Valencia, Spain.}
\author{F.~Galtarossa}
  \affiliation{Universit\'e Paris-Saclay, CNRS/IN2P3, IJCLab, 91405 Orsay, France}
\author{V.~Girard-Alcindor}
	\affiliation{Grand Acc\'el\'erateur National d'Ions Lourds (GANIL), CEA/DRF-CNRS/IN2P3, Caen, France}
\author{V.~Gonzales}
  \affiliation{Departamento de Ingenier\'ia Electr\'onica, Universitat de Valencia, Burjassot, Valencia, Spain}
\author{A.~Gottardo}
	\affiliation{INFN Laboratori Nazionali di Legnaro, Legnaro, Italy.}
\author{F.~Hammache}
  \affiliation{Universit\'e Paris-Saclay, CNRS/IN2P3, IJCLab, 91405 Orsay, France}
\author{L.J.~Harkness-Brennan}
  \affiliation{Oliver Lodge Laboratory, The University of Liverpool, Liverpool, UK.}
\author{H.~Hess}
  \affiliation{Institut für Kernphysik, Universität zu Köln, Zülpicher Str. 77, D-50937 Köln, Germany}
\author{D.S~Judson}
  \affiliation{Oliver Lodge Laboratory, The University of Liverpool, Liverpool, UK.}
\author{A.~Jungclaus}
  \affiliation{Instituto de Estructura de la Materia, CSIC, Madrid, E-28006 Madrid, Spain}
\author{A.~Kaşkaş}
  \affiliation{Department of Physics, Faculty of Science, Ankara University, 06100 Besevler - Ankara, Turkey}
\author{Y.H.~Kim}
	 \affiliation{Institue Laue-Langevin, Grenoble, France.}
\author{A.~Ku\c{s}o\u{g}lu}
	 \affiliation{Department of Physics, Faculty of Science, Istanbul University, Vezneciler/Fatih, Istanbul, Turkey.}
\author{M.~Labiche}
  \affiliation{STFC Daresbury Laboratory, Daresbury, Warrington, WA4 4AD, UK}
\author{S.~Leblond}
	\affiliation{Grand Acc\'el\'erateur National d'Ions Lourds (GANIL), CEA/DRF-CNRS/IN2P3, Caen, France}
\author{C.~Lenain}
     \affiliation{Universit\'e de Caen Normandie, ENSICAEN, CNRS/IN2P3, LPC Caen UMR6534, F-14000 Caen, France.}
\author{S.M.~Lenzi}
	 \affiliation{INFN, Sezione di Padova, I-35131 Padova, Italy.}
\author{S.~Leoni}
  \affiliation{INFN Sezione di Milano, I-20133 Milano, Italy}
\author{H.~Li}
	\affiliation{Grand Acc\'el\'erateur National d'Ions Lourds (GANIL), CEA/DRF-CNRS/IN2P3, Caen, France}  
\author{J.~Ljungvall}
  \affiliation{Universit\'e Paris-Saclay, CNRS/IN2P3, IJCLab, 91405 Orsay, France}
\author{J.~Lois-Fuentes}
	 \affiliation{IGFAE and Dpt. de Física de Partículas, Univ. of Santiago de Compostela, Santiago de Compostela, Spain}
\author{A.~Lopez-Martens}
  \affiliation{Universit\'e Paris-Saclay, CNRS/IN2P3, IJCLab, 91405 Orsay, France}
\author{A.~Maj}
  \affiliation{The Henryk Niewodnicza\'nski Institute of Nuclear Physics, Polish Academy of Sciences, 31-342 Krak\'ow, Poland}
\author{R.~Menegazzo}
	 \affiliation{INFN, Sezione di Padova, I-35131 Padova, Italy.}
\author{D.~Mengoni}
	\affiliation{Dipartimento di Fisica, Universit\`a di Padova, Padova, Italy.}
	 \affiliation{INFN, Sezione di Padova, I-35131 Padova, Italy.}
\author{C.~Michelagnoli}
	\affiliation{Grand Acc\'el\'erateur National d'Ions Lourds (GANIL), CEA/DRF-CNRS/IN2P3, Caen, France}
	 \affiliation{Institue Laue-Langevin, Grenoble, France.}
\author{B.~Million}
  \affiliation{INFN Sezione di Milano, I-20133 Milano, Italy}
\author{D.R.~Napoli}
  \affiliation{INFN Laboratori Nazionali di Legnaro, Legnaro, Italy.}
\author{J.~Nyberg}
  \affiliation{Department of Physics and Astronomy, Uppsala University, SE-75120 Uppsala, Sweden}
\author{G.~Pasqualato}
	\affiliation{Dipartimento di Fisica, Universit\`a di Padova, Padova, Italy.}
	 \affiliation{INFN, Sezione di Padova, I-35131 Padova, Italy.}
\author{Zs.~Podolyak}
  \affiliation{Department of Physics, University of Surrey, Guildford, GU2 7XH, UK}
\author{A.~Pullia}
  \affiliation{INFN Sezione di Milano, I-20133 Milano, Italy}
\author{B.~Quintana}
  \affiliation{Laboratorio de Radiaciones Ionizantes, Departamento de Física Fundamental, Universidad de Salamanca, E-37008 Salamanca, Spain}
\author{F.Recchia}
  \affiliation{Dipartimento di Fisica, Universit\`a di Padova, Padova, Italy.}
	 \affiliation{INFN, Sezione di Padova, I-35131 Padova, Italy.}
\author{D.~Regueira-Castro}
	 \affiliation{IGFAE and Dpt. de Física de Partículas, Univ. of Santiago de Compostela, Santiago de Compostela, Spain}
\author{P.~Reiter}
  \affiliation{Institut für Kernphysik, Universität zu Köln, Zülpicher Str. 77, D-50937 Köln, Germany}
\author{K.~Rezynkina}
	 \affiliation{Universit\'e de Strasbourg, CNRS, IPHC UMR 7178, F-67000 Strasbourg, France}
\author{J.S.~Rojo}
	 \affiliation{Department of Physics, University of York, York, UK.}
\author{M.D.~Salsac}
	 \affiliation{Irfu, CEA, Université Paris-Saclay, F-91191 Gif-sur-Yvette, France}
\author{E.~Sanchis}
  \affiliation{Departamento de Ingenier\'ia Electr\'onica, Universitat de Valencia, Burjassot, Valencia, Spain}
\author{M.~Şenyiğit}
  \affiliation{Department of Physics, Faculty of Science, Ankara University, 06100 Besevler - Ankara, Turkey}
\author{M.~Siciliano}
	 \affiliation{Irfu, CEA, Université Paris-Saclay, F-91191 Gif-sur-Yvette, France}
  \affiliation{Physics Division, Argonne National Laboratory, Lemont (IL), United States.}
\author{D.~Sohler}
  \affiliation{Institute for Nuclear Research, Atomki, 4001 Debrecen, Hungary}
\author{O.~Stezowski}
	 \affiliation{Université de Lyon, Université Lyon-1, CNRS/IN2P3, UMR5822, IP2I, F-69622 Villeurbanne Cedex, France}
\author{Ch.~Theisen}
	 \affiliation{Irfu, CEA, Université Paris-Saclay, F-91191 Gif-sur-Yvette, France}
\author{A.~Utepov}
	\affiliation{Grand Acc\'el\'erateur National d'Ions Lourds (GANIL), CEA/DRF-CNRS/IN2P3, Caen, France}
    \affiliation{Universit\'e de Caen Normandie, ENSICAEN, CNRS/IN2P3, LPC Caen UMR6534, F-14000 Caen, France.}
\author{J.J.~Valiente-Dob\'on}
    \affiliation{INFN Laboratori Nazionali di Legnaro, Legnaro, Italy.}
\author{D.~Verney}
  \affiliation{Universit\'e Paris-Saclay, CNRS/IN2P3, IJCLab, 91405 Orsay, France}
\author{M.~Zielinska}
	 \affiliation{Irfu, CEA, Université Paris-Saclay, F-91191 Gif-sur-Yvette, France}

    \begin{abstract}
The excited states of unstable $^{20}$O were investigated via $\gamma$-ray spectroscopy following the $^{19}$O$(d,p)^{20}$O reaction at 8 $A$MeV. 
By exploiting the Doppler Shift Attenuation Method, the lifetime of the 2$^+_2$ and 3$^+_1$ states were firmly established. 
From the $\gamma$-ray branching and E2/M1 mixing ratios for transitions deexciting the 2$^+_2$ and 3$^+_1$ states, the B(E2) and B(M1) were determined. Various chiral effective field theory Hamiltonians, describing the nuclear properties beyond ground states, along with a standard USDB interaction, were compared with the experimentally obtained data. 
Such a comparison for a large set of $\gamma$-ray transition probabilities with the valence space in medium similarity renormalization group \textit{ab-initio} calculations was performed for the first time in a nucleus far from stability.  
It was shown that the \textit{ab-initio} approaches using chiral EFT forces are challenged by detailed high-precision spectroscopic properties of nuclei. 
The reduced transition probabilities were found to be a very constraining test of the performance of the \textit{ab-initio} models. 

    \end{abstract}
    
    \maketitle
    
    \textit{Introduction}.\textemdash 
Nuclear structure studies aim at understanding the properties of atomic nuclei based on nucleons interacting in the nuclear medium by combined strong, electromagnetic and weak interactions. 
Chiral effective field theory (EFT) provides a framework for nuclear forces based on quantum chromodynamics which, together with \textit{ab-initio} many-body approaches, allows one to perform first-principle nuclear structure calculations including two- and three-nucleon forces in various regions of the Segr\'e chart~\cite{Hebeler2015,hergert2020,stroberg2021,hu2021}. 
Previous studies of neutron-rich isotopes have proven to be especially suitable to establish advanced theoretical calculations based on chiral EFT forces. 
In particular, the neutron drip line for oxygen presents a strong anomaly with $^{24}$O being the last bound isotope, whereas theoretical predictions positioned the drip line at doubly-magic $^{28}$O~\cite{janssens2009,kanungo2009,hoffman2008,ozawa2000,KONDO2023}. 
This puzzle was solved by the introduction of chiral EFT three-body forces~\cite{Otsuka2010}. 
These contributions have been studied extensively in subsequent works, especially in comparison with mass~\cite{michimasa2018, leistenschneider2021, Wienholtz2013, manea2020, mougeot2021,silwal2022}, charge radius~\cite{GarciaRuiz2016,deGroote2019, Kaufmann2020, Koszorus2020,MalbrunotEttenauer2021, Kaur2022, Sommer2022}, and electromagnetic moment~\cite{GarciaRuiz2015,Vernon2013, Bai2022} measurements of neutron-rich systems. 
The present challenge is to obtain unambiguous experimental measurement to compare to different \textit{ab-initio} calculations to improve their accuracy and predictive power. 
Electromagnetic transition probabilities play a major role in testing the quality of the chiral EFT interaction with \textit{ab-initio} approaches~\cite{Forssen2013,hu2021}, since they are connected to the nuclear wave functions. The comparison between high precision measurements in excited states and state-of-the-art \textit{ab-initio} calculations provides a sensitive probe of the nuclear structure details comparable to nuclear masses or charge radii.
The isotopic chain of oxygen was identified as an ideal laboratory to benchmark state-of-the-art \textit{ab-initio} theory \cite{Hebeler2015}. In neutron-rich oxygen, the introduction of three-body forces  induces a repulsion between the neutron 1$s_{1/2}$ and 0$d_{3/2}$ orbitals defining the drip-line~\cite{Otsuka2010}. 
Detailed spectroscopy of these orbitals at the drip-line remains an experimental challenge but relevant information can be obtained in slightly less exotic nuclei, such as $^{20}$O~\cite{Ciemala2020}. 
In this Letter, we present the spectroscopic study of non-yrast states in $^{20}$O using state-of-the-art instrumentation.\\

\textit{Experimental details}.\textemdash 
The $^{20}$O nucleus was populated in the $^{19}$O$(d,p)^{20}$O* direct reaction in inverse kinematics, using a pure radioactive beam post-accelerated to $8$ $A$MeV, with an average intensity of $4 \times 10^5$ pps,  delivered by the SPIRAL1 accelerator complex in GANIL and impinged on a deuterated polyethylene target (CD$_2$).  Two types of targets were employed in the experiment: a 0.3 mg/cm$^2$-thick self-supporting CD$_2$ target for spectroscopy measurements and a 0.3 mg/cm$^2$-thick target deposited on a 24.4 mg/cm$^2$-thick Au backing (hereinafter mentioned as CD$_2$ and CD$_2$+Au, respectively). 
Using the CD$_2$ target, detailed spectroscopy was performed and the CD$_2$+Au was used for measuring the lifetime of the populated excited states using the Doppler Shift Attenuation Method (DSAM)~\cite{Nolan_1979}. 
The measurements were performed in triple coincidence: the beam-like recoils were detected in the VAMOS++ magnetic spectrometer~\cite{Rejmund2011} to reject the background coming from fusion-evaporation and fusion-fission events, protons were measured at backward angles by the MUGAST array~\cite{Assie2021} and, at backward angles, the AGATA array~\cite{akkoyun2012} was employed for the detection of the $\gamma$~rays emitted by the excited nucleus.

The coupling of these three instruments provides a large solid angle for detection of recoiling nuclei, a high precision kinematic reconstruction and a unique sensitivity for $\gamma$ rays emitted in flight thanks to $\gamma$-ray Tracking Algorithms~\cite{Lopez2004}, resulting in unprecedented Doppler correction capabilities. 
This unique combination of direct reaction and state-of-the-art spectrometers allows one to perform a combined charged particle and  $\gamma$-ray spectroscopy, along with the measurement of  sub-picosecond lifetimes. 
In particular, the MUGAST array allows for the selection on an event-by-event basis of the excited states directly populated in the final $^{20}$O$^*$ nucleus and measure its velocity at the reaction moment for each of the populated states to perform a feeding-free, fully controlled and high accuracy lifetime measurement. 
More details on the experimental apparatus and data analysis procedure can be found in Ref.~\cite{Assie2021}.\\

\begin{figure}
\centering
\includegraphics[width=1.\columnwidth]{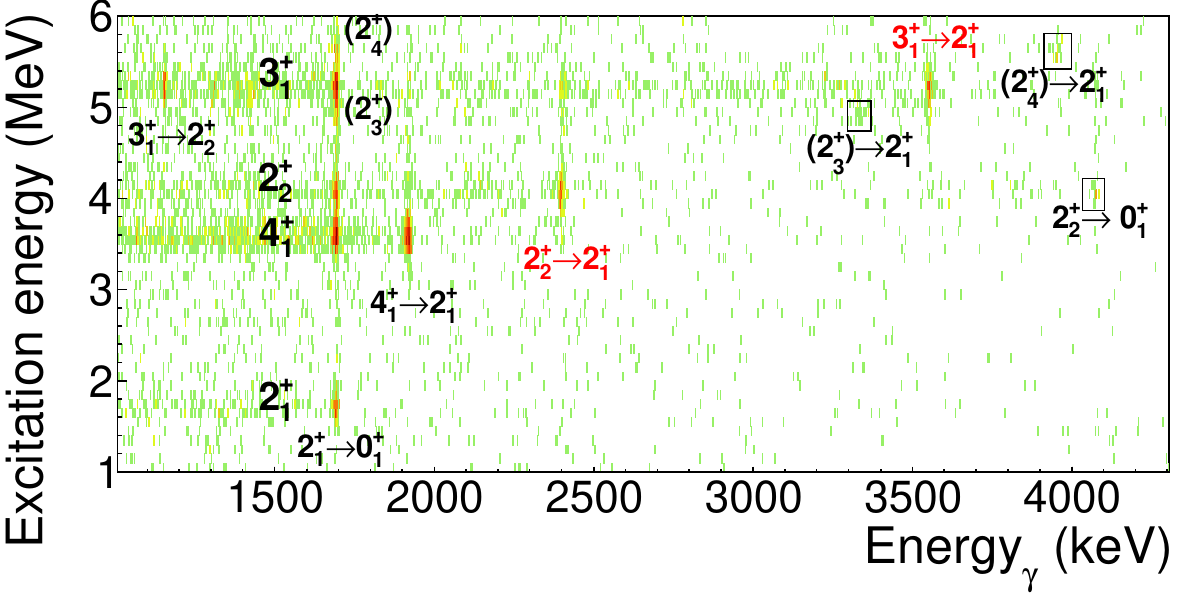}
\caption{(Color online) Two dimensional correlation between the $^{20}$O excitation energy on the y-axis and the corresponding $\gamma$-ray decays on the x-axis. The transitions from which the lifetimes are extracted are highlighted in red. The weakest transitions depopulating the 2$^+_{2,3,4}$ states are marked with black boxes.}
\label{fig:egammaEx}
\end{figure}

\textit{Spectroscopic study}.\textemdash
The energy and angle of the protons detected in MUGAST allowed for the reconstruction of the $^{20}$O excitation energy spectrum using NPTool~\cite{nptool,Assie2021} (see also the supplementary material).
The ground state, the $2^+_1$ at 1.67 MeV, the $4^+_1$ at 3.55 MeV, the $2^+_2$ at 4.07 MeV and $3^+_1$ states at 5.23 MeV were observed.  
Moreover, two additional states, already identified in~\cite{Hoffman2012} and both tentatively assigned as $J^{\pi}=(2^+_{3,4})$, were observed.  
Excited states above the neutron separation threshold in $^{20}$O were populated~\cite{ZanonPhD}.
The correlation between the excitation energy of $^{20}$O and the emitted $\gamma$-rays is shown in Fig.~\ref{fig:egammaEx}. 
The transitions used in the line-shape analysis for the lifetime extraction are highlighted in red in the figure.

\begin{figure}
\centering
\includegraphics[width=1.\columnwidth]{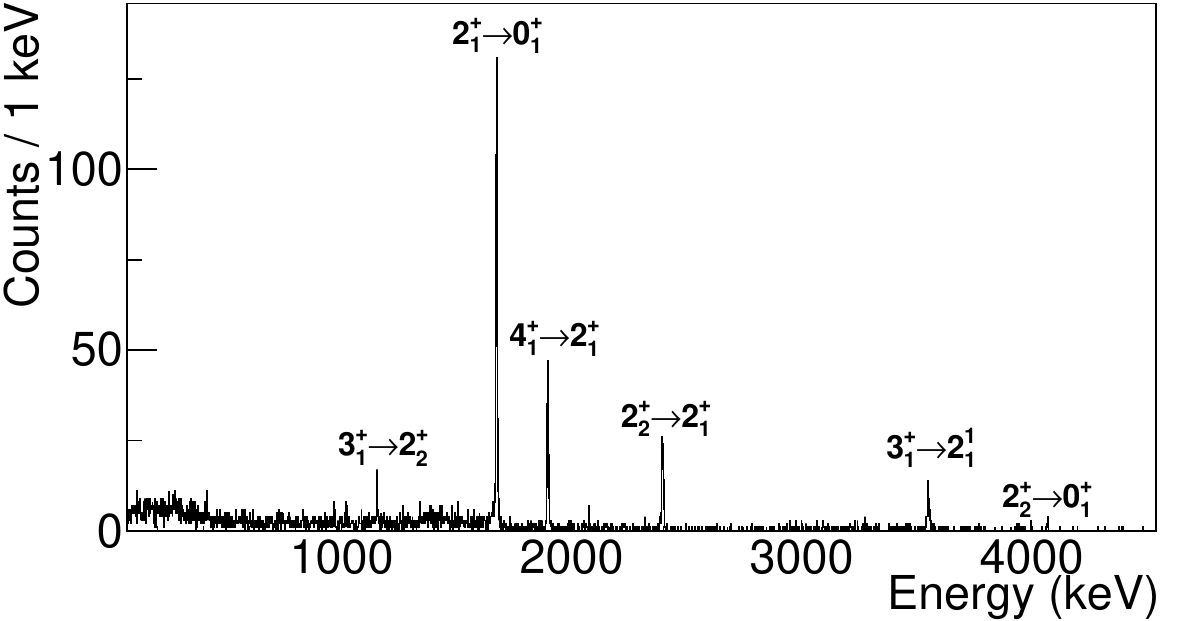}
\caption{Prompt $\gamma$-ray spectrum following the $^{19}$O$(d,p)^{20}$O reaction measured in AGATA in coincidence with MUGAST and VAMOS++.}
\label{fig:egamma}
\end{figure}

The prompt Doppler-corrected tracked $\gamma$-ray spectrum measured in AGATA in triple coincidence using the CD$_2$ target is shown in Fig.~\ref{fig:egamma}. The decays of the 2$^+_1$, 4$^+_1$, 2$^+_2$ and 3$^+_1$ excited states are clearly visible.
The reconstruction of the level scheme obtained via particle-$\gamma$ spectroscopy combining MUGAST and AGATA is reported in detail in ~\cite{Zanon2021,Zanon2022,ZanonPhD}. 
The complete spectroscopic information is summarized in the supplementary material. 
The branching ratios of the $2^+_2 \to 2^+_1$ and $2^+_2 \to 0^+_1$ transitions were measured to be 0.88(1) and 0.12(1), respectively. 
For the $3^+_1 \to 2^+_2$ and $3^+_1 \to 2^+_1$ transitions, the measured branching ratios were 0.28(1) and 0.72(1), respectively. \\ 

\textit{Lifetime measurements}.\textemdash
Previous experiments provided the lifetime measurements of the 2$^{+}_{1}$ ($\tau$=10.5(4)~ps \cite{RAGHAVAN1989189}) and 2$^{+}_{2}$ ($\tau$=150$^{+80}_{-30}$~fs \cite{Ciemala2020}) states. In the present work, lifetimes were extracted by fitting the line-shape of the transitions in the $\gamma$-ray spectra of the CD$_2$+Au dataset with realistic Monte Carlo simulations. 
The simulations have been performed using the AGATA Geant4 simulation code~\cite{Farnea2010}, that includes the geometry of the array and the reaction event with the emission of the beam- and target-like particles as well as $\gamma$ rays. 
Line-shape analysis based on Monte Carlo simulations for AGATA was already performed in the fs range~\cite{Ciemala2021, Ziliani2020,MichelagnoliPhD}, proving the capabilities of the apparatus in this range of lifetimes.

The simulation has been optimized by adjusting the parameters corresponding to the response function of the detectors at the time of the experiment to reduce the sources of systematic errors.
The velocity distributions at reaction point were measured and used as an input of the simulation of the decay for each of the investigated states.
The reproduction of the energy loss in the target and degrader was tested for both CD$_2$ and CD$_2$+Au datasets on the $2^+_1 \to 0^+_1$ transition.
The energy and Full-Width Half Maximum of the transition were in agreement within the detector resolution.
The comparison is provided in the supplementary material.

The lifetime of the $2^+_2$ state was extracted by fitting the line-shape of the simulated $2^+_2 \to 2^+_1$ transition to the experimental one, obtained by requiring the coincidence with the $2^+_2$ state in the excitation energy and thus removing the influence of feeders.
The simulations were performed varying two parameters: the energy of the transition at rest and the lifetime of the $2^+_2$ state. 
The minimum was attested at 70(10)~fs (70(14)~fs for 90\% confidence).
\begin{figure}
\includegraphics[width=0.45\textwidth]{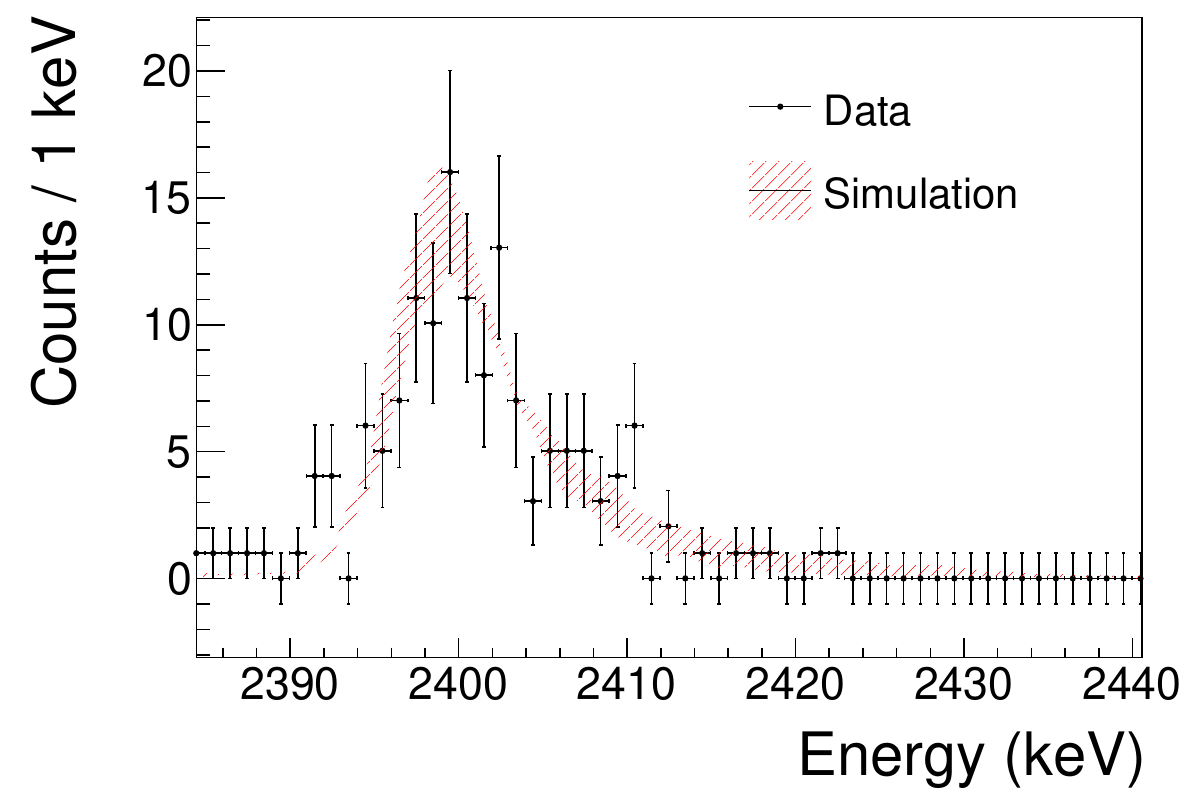}
\caption{(Color online) Comparison between experimental data and simulation for the lifetime of the $2^+_2$ state in $^{20}$O.}
\label{fig:2p2}
\end{figure}
In Fig.~\ref{fig:2p2}, the red hatched area represents the simulation output within the $1\sigma$ limit. The two dimensional $\chi^2$ surface is shown in the supplementary material. 
To extract the lifetime of the $3^+_1$ state, the $3^+_1 \to 2^+_1$ transition at 3552.6 keV was chosen.
Similarly to the procedure for the $2^+_2$ state, the simulated spectrum was fitted to the experimental one after gating on the excitation energy of the $3^+_1$ state.
Using the least-$\chi^2$ procedure, a lifetime of $\tau$ = 54(9)~fs was obtained (54(12)~fs for 90\% confidence).\\

    \textit{Discussion}.\textemdash 
The present measurement confirms the short lifetime of the 2$^+_2$ state and the conclusion drawn in \cite{Ciemala2020}  in spite of a tension between the results of the two experiments. 
The origin of the discrepancy is not completely understood. The computation of a longer lifetime can arise from systematical effects in the initial velocity distribution model or unobserved side-feeding contribution. Such systematic errors are cancelled in our improved experimental approach. 
The use of the $(d,p)$ reaction combined to a thin target layer where $^{20}$O is produced, followed by an optimized gold thick foil to develop the DSAM profile, allows one to determine, on an event-by-event basis, the entry point in the recoil from the measured excitation energy and the initial velocity distribution for each state used to extract the lifetimes. 
The reduced transition probabilities have been extracted from the measured transition energies, branching ratios and lifetimes, and reported in Table~\ref{tab:be2}. 
A measured value of the mixing ratio $\delta(E2/M1)$ for the $2^+_2 \to 2^+_1$ transition reported earlier is $-0.18(8)$~\cite{Young1981}.  The mixing ratios of the $3^+_1 \to 2^+_{1,2}$ transitions are experimentally unknown, preventing us to extract model independent transition probabilities. \\

The present experimental results have been compared to \textit{ab-initio} calculations using the valence-space in-medium similarity renormalization group (VS-IMSRG)~\cite{Stroberg2019}. 
The calculations were performed at the VS-IMSRG(2) level, building an effective shell-model Hamiltonian for $^{20}$O in the 0$d_{5/2}$, 1$s_{1/2}$, 0$d_{3/2}$ configuration space for protons and neutrons. 
The $E2$ and $M1$ transition operators were evolved consistently with the VS-IMSRG keeping up to two-body operators, but meson-exchange currents, explored for $M1$ transitions in very light nuclei~\cite{Marcucci2008,Pastore2013,Friman-Gayer2021}, were not included.
As a starting point, three well established nuclear Hamiltonians based on chiral EFT with three-nucleon forces were used: (i) 1.8/2.0(EM)~\cite{Hebeler2011,Simonis2017}, which reproduces very well ground-state energies up to heavy nuclei~\cite{stroberg2021,Miyagi2022} and was used for $^{20}$O in Ref.~\cite{Ciemala2020}; (ii) the more recent N$^3$LO$_{lnl}$~\cite{Soma2020} and (iii) N$^2$LO$_{GO}$~\cite{Jiang2020}, which includes explicit $\Delta$-isobar degrees of freedom. 
In addition, standard shell-model calculations using the USDB interaction~\cite{brown2006} in the same configuration space were performed. 
Unlike in  the VS-IMSRG calculations, the bare $M1$ and $E2$ operators and therefore neutron effective charges $e_n=0.4$~\cite{heil2020}, were used along with the USDB interaction. For the VS-IMSRG and configuration-interaction calculations the codes imsrg++~\cite{imsrg++} and KSHELL~\cite{Shimizu:2019xcd} were used, respectively.

\begin{figure*}
    \centering
    \includegraphics[width=0.9\textwidth]{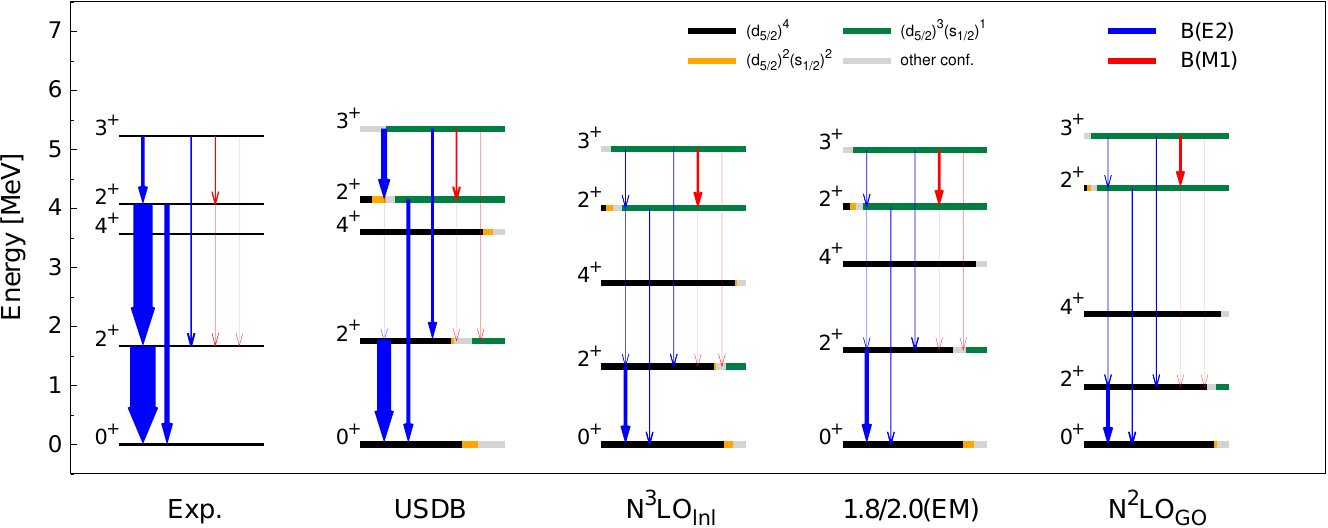}
    \caption{(Color online) Experimental $^{20}$O excited states compared to theoretical USDB shell-model calculations and VS-IMSRG results obtained with three different Hamiltonians. For theoretical states, the leading configurations are reported with a color code for each level: $(d_{5/2})^4$ (black), $(d_{5/2})^3(s_{1/2})^1$ (green) and $(d_{5/2})^2(s_{1/2})^2$ (violet). Other configurations are shown in gray. The colored bars length is proportional to each contribution in the wave-function. The measured and calculated $B(E2)$s (blue) and $B(M1)$s (red) are also reported proportionally to the arrows thickness. }
    \label{fig:spectroscopy}
\end{figure*}

Fig.~\ref{fig:spectroscopy} compares the experimental low-lying excitation spectra of $^{20}$O with the results of the theoretical calculations. 
Additionaly, the experimental and calculated level scheme of $^{19}$O are also shown in the supplementary material. 
The excitation energies obtained in the \textit{ab-initio} approaches and the shell model are in general in good agreement with experiment, within hundreds of keV. 
The 1.8/2.0(EM) and USDB results are in the best agreement with the data. 
We emphasize the good agreement for the 1/2$^+$ excited state corresponding to the $(d_{5/2})^2(s_{1/2})^1$ configuration in $^{19}$O. 
Consistently, the evolution of its excitation energy, for different chiral EFT Hamiltonians, is correlated to the $2^+_1\rightarrow2^+_2$ energy difference in $^{20}$O (see supplementary material). 
It should be noted that there was no nuclear structure information on excitation energies \textit{etc.} in oxygen or similar systems used for the derivation of the chiral EFT Hamiltonians in \textit{ab-initio} approaches, while the shell model USDB interaction resulted from the fit to the selected nuclear structure data. 
The calculated wave functions of the states observe two main structures: the $0^+_1$, $2^+_1$ and $4^+_1$ yrast states are mainly due to the neutron $(0d_{5/2})^4$ configuration, while the $2^+_2$ and $3^+_1$ states are dominated by the $(0d_{5/2})^3(1s_{1/2})^1$ configuration. 
The color code on the level shows the amplitude of the main configurations ($\geq$ 10\%). 
The \textit{ab-initio} and shell-model calculations are in good agreement, but the shell model suggests more fragmented wave functions.

\begin{table*}
\centering
\begin{tabularx}{0.9\textwidth} { >{\centering\arraybackslash}X  >{\centering\arraybackslash}X  >{\centering\arraybackslash}X >{\centering\arraybackslash}X >{\centering\arraybackslash}X >{\centering\arraybackslash}X}
\hline
\hline											
	&	Exp.	&	USDB	&	N$^3$LO$_{lnl}$	&	1.8/2.0(EM)	&	$N^2$LO$_{GO}$	\\ [0.5ex]
\hline											
$B(E2;2^+_1 \to 0^+_1)$	&	$5.9(2)$	&	3.25	&	0.79	&	0.89	&	0.80	\\ [0.5ex]
\hline
$B(E2;2^+_2 \to 0^+_1)$	&	$1.3(2)$	&	0.77	&	0.21	&	0.20	&	0.26	\\ [0.5ex]
$B(E2;2^+_2 \to 2^+_1)$	&	$4(2)$	&	0.0005	&	0.089	&	0.070	&	0.18	\\ [0.5ex]
$B(M1;2^+_2 \to 2^+_1)$	&	$0.05(2)$	&	0.019	&	0.014	&	0.017	&	0.012	\\ [0.5ex]
\hline
$B(E2;3^+_1 \to 2^+_1)$	&	$0.32(7)$	&	0.57	&	0.16	&	0.17	&	0.17	\\ [0.5ex]
$B(M1;3^+_1 \to 2^+_1)$	&	$0.016(4)$ 	&	0.029	&	0.023	&	0.028	&	0.0089	\\ [0.5ex]
$B(E2;3^+_1 \to 2^+_2)$	&	$0.7(2)$	&	1.24	&	0.14	&	0.15	&	0.11	\\ [0.5ex]
$B(M1;3^+_1 \to 2^+_2)$	&	$0.19(4)$	&	0.32	&	0.53	&	0.55	&	0.56	\\ [0.5ex]
\hline
\hline
Binding energy& -23.74 \cite{Wang_2021} & -23.63 & -19.67 & -20.51& -22.71\\ [0.5ex]
\hline
\end{tabularx}
\caption{Comparison between experimental, shell-model (USDB) and \textit{ab-initio} (N$^3$LO$_{lnl}$, 1.8/2.0(EM), $N^2$LO$_{GO}$) transition probabilities. The $B(E2)$s are given in $e^2\mathrm{fm^4}$ and the $B(M1)$s in $\mu_N^2$. The experimental transition probabilities of the $3^+_1 \to 2^+_1$ ($3^+_1 \to 2^+_2$) have been calculated using a theoretical mixing ratio of $\delta = 0.13$ ($\delta = 0.019$), obtained from USDB. The binding energy with respect to  $^{16}$O is presented in MeV.\label{tab:be2}}
\end{table*}

The mixing ratio of transitions from the 3$^+_1$ state are experimentally unknown. For the following discussion, the theoretical values obtained using the USDB~\cite{brown2006} interaction were used to obtain the corresponding $B(E2)$ and $B(M1)$. 
The experimental and theoretical reduced transition probabilities are presented in Table~\ref{tab:be2} and reported in Fig.~\ref{fig:spectroscopy}. 
The $B(E2)$ reduced transition probabilities between the 2$^+_2$ state and the 0$^{+}_{1}$ and 2$^{+}_{1}$ states, and from the 3$^+_1$ state, were found experimentally small, consistently with their single-particle character, interpreted as $(0d_{5/2})^{4}\rightarrow(0d_{5/2})^{3}(1s_{1/2})^{1}$ single-particle transition.

The $B(E2)$ values are systematically underestimated in the \textit{ab-initio} calculations, as already observed in Ref.~\cite{heil2020} for $^{21}$O and discussed first in Ref.~\cite{parzuchowski2017} and recently in much detail in Ref.~\cite{stroberg2022,PhysRevC.105.034332}. 
The likely reason is the restriction to the VS-IMSRG(2) level, which leaves many-particle--many-hole correlations out of the evolved VS-IMSRG operator. 
With a neutron effective charge $e_n=0.4$, USDB $B(E2)$ results present a good agreement with experimental values, in particular, those involving the $3^+_1$ state. 
An even better agreement for some transitions is observed in USDB calculations using an effective charge $e_n=0.5$. 
However, high-precision spectroscopy data reveal that simple effective charges in the shell model seem not to reproduce all transitions simultaneously (see also the discussion in Ref.~\cite{heil2020}) and a more sophisticated treatment would be desirable. 
In addition, the $B(E2;2^+_2 \to 2^+_1)$ is underestimated by orders of magnitude in all the models, which suggests some deficiency of the wave functions. 
It could be related to the limited configuration space or to the insufficient configuration mixing. The latter is consistent with the results of the measurements of the spectroscopic factor ($S$) by~\citet{Hoffman2012}. 
The cross sections for 0$^+_1$ and 4$^+_1$ states were computed with $L=2$ transfer and large $S$-factor obtained are compatible with the occupancy of the $d$-orbitals only. 
Similarly, the cross section for 2$^+_2$ state is dominated by $L=0$ transfer consistent with the single-particle excitation into the 1$s_{1/2}$ neutron orbital. In contrast, the cross section for 2$^+_1$ state was obtained with large contributions of $L=0$ ($S=0.19$) and $L=2$ transfers ($S=0.43$) corresponding to a more fragmented wave function of the 2$^+_1$ state.

The theoretical $B(M1)$ reduced transition probabilities for the $2^+_2 \to 2^+_1$ transition are about a factor $3$ smaller than those obtained in the experiment, with a small difference between USDB and 1.8/2.0(EM) and lower values for the other two VS-IMSRG calculations. The $B(M1)$ for the $3^+_1$ state are in reasonably good agreement with the experiment, especially for USDB, where there is agreement within $(1-2)\sigma$. The three chiral EFT Hamiltonians reproduce well the $3^+_1 \rightarrow 2^+_1$ $B(M1)$ value, but they overestimate the $3^+_1 \rightarrow 2^+_2$ reduced transition probability by about a factor 2. Overall, for the $M1$ transitions there is a better agreement for the phenomenological USDB interaction. The \textit{ab-initio} results are of similar quality, with a slight preference for 1.8/2.0(EM) over the other two chiral EFT Hamiltonians. The agreement may improve when including meson exchange currents.  Hence, the measurements of the $B(M1)$ reduced transition probabilities appear to be very pertinent for testing {\it ab-initio} calculations based on chiral EFT Hamiltonians. It should be noted that the short lifetime measured in this work (lower than 100 fs) for the 2$^+_2$ state is incompatible with having at the same time a low B(E2) and a low B(M1), in the range of the theoretical predictions, for the $2^+_2\rightarrow2^+_1$ transition. 

    \textit{Conclusion}. \textemdash 
The lifetimes of the 2$^+_2$ and 3$^+_1$ excited states in $^{20}$O were measured by means of the DSAM technique via the direct $(d,p)$ reaction in inverse kinematics using a radioactive post-accelerated beam of $^{19}$O. 
A feeding-free lifetime for the 2$^+_2$ and the 3$^+_1$ states was extracted. For the first time in the key isotopic chain of oxygen, all spectroscopic observable obtained for yrast and non-yrast excited states in the neutron rich $^{20}$O were compared simultaneously to the results of \textit{ab-initio} calculations using chiral EFT forces and provide the results in reasonable agreement with the experimental data. 
The reduced transition probabilities, $B(M1)$ and $B(E2)$ in particular, provide a very constraining test of the performance of the \textit{ab-initio} models.  
Many improvements in \textit{ab-initio} calculations are still to be envisaged  like including meson exchange currents or many-particle--many-hole correlations by releasing the restriction to the VS-IMSRG(2) level, so that the predictive power can reach and exceed that of the conventional phenomenological shell-model approaches. 
This work paves the way for lifetime measurements in exotic nuclei using next-generation radioactive beam facilities under construction worldwide to be compared with state of the art \textit{ab-initio} calculations. 

\textit{Data Availability Statements}. \textemdash 
The supporting data for this article are from the e775s experiment and are registered as https://doi.org:10.10.26143/GANIL-2020-E775S following the GANIL Data Policy.

\begin{acknowledgments}
We acknowledge the GANIL facility for provision of heavy-ion beams and we would like to thank J. Goupil, G. Fremont, L. M\'enager, A. Giret for assistance in using the G1 beam line and its instrumentation. This work was supported by STFC(UK). This research was also supported by the OASIS Project No. ANR-17-CE31-0026, by Project PRIN 2017P8KMFT - CUP I24I19000140001, by the U.S. Department of Energy, Office of Science, Office of Nuclear Physics, under contract number DE-AC02-06CH11357, by the European Regional Development Fund with Contract No. GINOP-2.3.3-15-2016-00034 and by the MCIN/AEI /10.13039/501100011033, Spain, grants PID2020-118265GB-C41, PID2020-118265GB-C42, by Generalitat Valenciana, Spain, grants PROMETEO/2019/005, CIAPOS/2021/114 and by the FEDER EU funds, by the Deutsche Forschungsgemeinschaft (DFG, German Research Foundation) -- Project-ID 279384907 -- SFB 1245, and by the European Research Council (ERC) under the European Union’s Horizon 2020 research and innovation program (grant agreement No 101020842). 
Cloud Veneto is acknowledged for the use of computing
and storage facilities~\cite{cloudveneto}. 
The nucleon-nucleon and three-nucleon interaction matrix elements used in the VS-IMSRG calculations are generated by NuHamil \cite{Miyagi2023}.
\\
\end{acknowledgments}
 
\bibliographystyle{apsrev4-1}
\bibliography{main.bib}

\end{document}